\documentstyle [11pt,epsf]{article}
\begin{document}
\begin{flushright}
PREPRINT \\
INR-P0989/98\\
JINR E2-98-265\\
US-FT/17-98\\
SEPTEMBER 1998
\end{flushright}

\begin{center}

{\Large \bf
The QCD analysis of the CCFR data for $xF_3$:
higher twists and $\alpha_s(M_Z)$ extractions at the NNLO and
beyond}

\vspace{0.1cm}

{\bf A.L. Kataev}

\vspace{0.1cm}

{\baselineskip=14pt Institute for Nuclear Research
of the Academy of Sciences of Russia,\\ 117312 Moscow, Russia},

\vspace{0.1cm}

{\bf G. Parente}

\vspace{0.1cm}

{\baselineskip=14pt Department of Particle Physics, University
of Santiago de Compostela,\\ 15706 Santiago de Compostela, Spain}

\vspace{0.1cm}

{\bf  A.V. Sidorov}
\vspace{0.1cm}

{\baselineskip=14pt Bogoliubov Laboratory of Theoretical Physics,\\ Joint
Institute for Nuclear Research,\\ 141980 Dubna, Russia}
\end{center}
\abstract
{The improved next-to-next-to-leading order (NNLO) QCD
analysis of the experimental data of the CCFR
collaboration for the $xF_3$ structure function is made.
Theoretical ambiguities of the NNLO fits are estimated
by means of the Pad\'e resummation technique, which was  applied both
in the expanded and non-expanded forms.
The
NNLO and the new N$^3$LO $\alpha_s(Q^2)$ $\overline{MS}$-matching
conditions are used.
In the process of the fits we are taking into account the
target mass corrections and the twist-4 $1/Q^2$-terms.
Our  NNLO results for $\alpha_s(M_Z)$ values,
extracted from the CCFR $xF_3$ data, are
$\alpha_s(M_Z)=
0.118 \pm 0.002 (stat)\pm 0.005 (syst) \pm 0.003(theory)$
provided the twist-4
contributions are fixed through the infrared renormalon model and
$\alpha_s(M_Z)=0.116^{+ 0.006}_{-0.007}(stat)\pm 0.005 (syst) \pm 0.004(theory)$
provided the twist-4 terms are considered as the free parameters.
It is shown that the extracted at the NNLO order $x$-shape
of the twist-4 correction is almost unchanged  after
application of
the  Pad\'e resummation approximant.
}

\newpage
{\bf 1. Introduction.}~

Deep-inelastic lepton-nucleon scattering (DIS) belongs to the
classical and continuously studying processes in the modern
particle physics. The traditionally measurable characteristics
of $\nu N$ DIS are the SFs $F_2$ and $xF_3$.
It should be stressed that the
program of getting the information about the behavior of the
SFs of $\nu N$ DIS is among the aims of the experimental
program of Fermilab Tevatron and CCFR/NuTeV collaboration
in particular. The CCFRR collaboration
started to study the $\nu N$ scattering process over 1980 year
\cite{CCFRR}. The data for
the SFs of $\nu N$ DIS, obtained by the
follower of the CCFRR collaboration,
namely CCFR group, was distributed among the potential users in the
beginning of 1997 \cite{Seligman}, while the final results of the original
CCFR DGLAP \cite{DGLAP} NLO analysis of this data was presented
in the journal publication of Ref.\cite{Seligman}.

This experimental information was already used in the process
of different NLO analysis, performed by CITEQ,
MRST and GRV  groups (see Refs.\cite{CITEQ,MRST,GRV} correspondingly).
The subsequent steps of performing NLO and first NNLO analysis
of  the CCFR data with the help
of the Jacobi polynomial - Mellin moments version of the DGLAP
method
were made in Refs.\cite{KS}-\cite{KKPS3}
(the definite stages in the development of this formalism
are described in Refs.\cite{PS}-\cite{Kretal}).

In the process of the analysis of Refs.\cite{KKPS1}-\cite{KKPS3}
the authors used the information about the NNLO corrections to the coefficient
functions \cite{WZ} and the available analytical expressions
for the NNLO corrections to the anomalous dimensions
of the NS moments with $n=2,4,6,8,10$ \cite{LRV1}, supplemented
with the given in Ref.\cite{KKPS1} $n=3,5,7,9$ similar numbers,
obtained using the smooth interpolation procedure, which was previously
proposed in Ref.\cite{PKK}. Moreover, the attempts to extract
the shape of the twist-4 contributions and study the predictive
abilities of the IR-renormalon (IRR) model of Ref.\cite{DW} were performed
(for the definite details of modeling the effects of the power-suppressed
contributions
to measurable physical quantities using the IRR
language see  Ref.\cite{IRR} and the review of Ref.\cite{Beneke}).

However,  the important question of the estimation of
theoretical uncertainties of the NNLO analysis of the CCFR data
of Ref.\cite{KKPS2} was still non-analyzed in detail.
These uncertainties are determined by

1) the differences in the definitions of  $\alpha_s(Q^2)$
matching conditions (see e.g. \cite{BW, Marciano, DV}) which are responsible
for penetrating into the energy region, characteristic for $f=5$ numbers
of flavours, where the pole of the $Z^0$-boson is manifesting itself;

2) the incorporation into the condition of Ref.\cite{BW} recalculated
NNLO QCD corrections \cite{LRVS} and the newly
calculated N$^3$LO corrections, namely the 4-loop coefficient
of the QCD $\beta$-function \cite{RVL} and the N$^3$LO-term
\cite{CKS} in the matching condition of Ref.\cite{BW};

3) the consideration  of theoretical
uncertainties due to other non-calculated N$^3$LO contributions into
the coefficient functions and the anomalous dimensions function.


This work is devoted to the analysis of the important problems
outlined above and to the more detailed extraction of the values
of $\alpha_s(M_Z)$ and the $x$-shape of the
twist-4 power-suppressed term at available orders of perturbative QCD with
taking into account the effects
enumerated above. We are supplementing the NNLO fits of
Ref.\cite{KKPS2} by the N$^3$LO analysis, which is based on the
application of the Pad\'e resummation technique (for the review see
Ref.\cite{Pade}), developed in QCD in the
definite form in Refs.\cite{SEK,EGKS} and considered previously as
the possible method of fixing theoretical uncertainties in the analysis of
DIS data in Ref.\cite{Sid}. It should be stressed that
{\it a posteriori} this technique gives the results similar  to
those, obtained with the the  help of the application
of different methods of fixing scale-scheme
dependence ambiguities (compare the results of Ref.\cite{KatSt} with the results
of Refs.\cite{SEK,EGKS} obtained using  the Pad\'e resummation technique).
Thus, our analysis could be considered as the attempt to estimate
perturbative QCD uncertainties beyond the NNLO level. Moreover, it could give
us the hint whether the outcomes of the NNLO fits, related to perturbative
and non-perturbative sectors, stay stable after the inclusion of the
explicitly calculated and estimated N$^3$LO QCD corrections.

{\bf 2. The theoretical background of the QCD analysis.}~

Let us  define the Mellin moments
for the NS SF $xF_3(x,Q^2)$:
\begin{equation}
M_n^{NS}(Q^2)=\int_0^1 x^{n-1}F_3(x,Q^2)dx
\end{equation}
where $n=2,3,4,...$. The theoretical expression
for these moments
obey the following renormalization group equation
\begin{equation}
\bigg(\mu\frac{\partial}{\partial\mu}+\beta(A_s)\frac{\partial}
{\partial A_s}
+\gamma_{NS}^{(n)}(A_s)\bigg)
M_n^{NS}(Q^2/\mu^2,A_s(\mu^2))=0
\label{rg}
\end{equation}
where $A_s=\alpha_s/(4\pi)$.
The renormalization group functions are defined as
\begin{eqnarray}
\mu\frac{\partial A_s}{\partial\mu}=\beta(A_s)=-2\sum_{i\geq 0}
\beta_i A_s^{i+2}~~\nonumber \\
\mu\frac{\partial ln
Z_n^{NS}}{\partial\mu}=\gamma_{NS}^{(n)}(A_s) =\sum_{i\geq 0}
\gamma_{NS}^{(i)}(n) A_s^{i+1}
\end{eqnarray}
where
$ Z_n^{NS} $
are the renormalization constants of the corresponding
NS operators.  The solution of the renormalization group
equation
can be presented in the following form :
\begin{equation}
\frac{ M_{n}^{NS}(Q^2)}{M_{n}^{NS}(Q_0^2)}=
exp\bigg[-\int_{A_s(Q_0^2)}^{A_s(Q^2)}
\frac
{\gamma_{NS}^{(n)}(x)}{\beta(x)}dx\bigg]
\frac{C_{NS}^{(n)}(A_s(Q^2))}
{C_{NS}^{(n)}(A_s(Q_0^2))}
\label{mom}
\end{equation}
where $M_n^{NS}(Q_0^2)$ is the phenomenological quantity related to the
factorization scale dependent factor.
It can be parametrized through the parton distributions at fixed momentum
transfer $Q_0^2$ as
\begin{equation}
M_n^{NS}(Q_0^2)=\int_0^1 x^{n-2} A(Q_0^2)x^{b(Q_0^2)}(1-x)^{c(Q_0^2)}
(1+\gamma(Q_0^2)x)dx
\end{equation}
with  $\gamma \neq 0$ or $\gamma=0$. In principle, following the
models of parton distributions, used in Refs.\cite{MRST,GRV},
one can add in the used  model for the SF the term, proportional to
$\sqrt{x}$. However, since this term is important in the region of
rather small $x$, we will neglect it in this  our analysis.

At the N$^3$LO the  expression for
the coefficient function $C_{NS}^{(n)}$  can be presented as
\begin{equation}
C_{NS}^{(n)}(A_s)=1+C^{(1)}(n)A_s+C^{(2)}(n)A_s^2+C^{(3)}(n)A_s^3,
\end {equation}
while the corresponding expansion of the anomalous dimensions term
is
\begin{equation}
exp\bigg[-\int^{A_s(Q^2)}
\frac{\gamma_{NS}^{(n)}(x)}{\beta(x)}dx\bigg]=
\big(A_s(Q^2)\big)^{\gamma_{NS}^{(0)}(n)/2\beta_0}
\times AD(n_s)
\end{equation}
where
\begin{equation}
AD(n_s)=
[1+p(n)A_s(Q^2)
+q(n)A_s(Q^2)^2+r(n)A_s(Q^2)^3]
\label{an}
\end{equation}
and  $p(n)$, $q(n)$ and $r(n)$ have the following form:
\begin{equation}
p(n)=\frac{1}{2}\bigg(\frac{\gamma_{NS}^{(1)}(n)}{\beta_1}-
\frac{\gamma_{NS}^{(0)}(n)}{\beta_0}\bigg)\frac{\beta_1}{\beta_0}
\end{equation}
\begin{equation}
q(n)=\frac{1}{4}\bigg( 2p(n)^2+
\frac{\gamma_{NS}^{(2)}(n)}{\beta_0}+\gamma_{NS}^{(0)}(n)
\frac{(\beta_1^2-\beta_2\beta_0)}{\beta_0^3}-\gamma_{NS}^{(1)}(n)
\frac{\beta_1}{\beta_0^2}\bigg)
\end{equation}
\begin{eqnarray}
r(n)=\frac{1}{6}\bigg(p(n)^3+6p(n)q(n)+
\frac{\gamma_{NS}^{(3)}(n)}{\beta_0}-\frac{\beta_1\gamma_{NS}^{(2)}(n)}
{\beta_0^2} \\ \nonumber
-\frac{\beta_2\gamma_{NS}^{(1)}(n)}{\beta_0^2}+
\frac{\beta_1^2\gamma_{NS}^{(1)}(n)}{\beta_0^3}
-\frac{\beta_1^3\gamma_{NS}^{(0)}(n)}{\beta_0^4}
-\frac{\beta_3\gamma_{NS}^{(0)}(n)}{\beta_0^2}+
\frac{2\beta_1\beta_2\gamma_{NS}^{(0)}(n)}{\beta_0^3}\bigg)
\end{eqnarray}
The coupling constant $A_s(Q^2)$ can be expressed
in terms
of the inverse powers of
$L=\ln(Q^2/\Lambda_{\overline{MS}}^2)$
as $A_s^{NLO}=A_s^{LO}+\Delta A_s^{NLO}$,
$A_s^{NNLO}=A_s^{NLO}+\Delta A_s^{NNLO}$ and
$A_s^{N^3LO}=A_s^{NNLO}+\Delta A_s^{N^3LO}$, where
\begin{eqnarray}
A_s^{LO}&=&\frac{1}{\beta_0
L} \\ \nonumber
\Delta A_s^{NLO}&=&
-\frac{\beta_1 ln(L)}{\beta_0^3 L^2}
\end{eqnarray}
\begin{equation}
\Delta A_s^{NNLO}=\frac{1}{\beta_0^5 L^3}[\beta_1^2 ln^2 (L)
-\beta_1^2 ln(L) +\beta_2\beta_0-\beta_1^2]
\end{equation}
\begin{eqnarray}
\Delta A_s^{N^3LO}=\frac{1}{\beta_0^7 L^4}[\beta_1^3 (-ln^3 (L)
+\frac{5}{2}ln^2 (L)
+2ln(L)-\frac{1}{2})
\\ \nonumber
-3\beta_0\beta_1\beta_2 ln(L)
+\beta_0^2\frac{\beta_3}{2}]~.
\end{eqnarray}

Notice that in our normalization the numerical expressions for
$\beta_0$, $\beta_1$, $\beta_2$ and $\beta_3$ read
\begin{eqnarray}
\beta_0&=&11-0.6667f \nonumber \\
\beta_1&=&102-12.6667f \nonumber \\
\beta_2&=&1428.50-279.611f+6.01852f^2 \nonumber \\
\beta_3&=&29243.0-6946.30f+405.089f^2+1.49931f^3
\end{eqnarray}
where the expression for $\beta_3$ was obtained in
Ref.\cite{RVL}. The inverse-log expansion for
$\Delta A_s^{N^3LO}$, which incorporates the information
about the coefficient $\beta_3$, was presented in Ref.\cite{CKS}.

Few words ought to be said about the used approximation for the
anomalous dimension function $\gamma_{NS}^{(n)}(A_s)$.
The analytical expression for its one-loop coefficient
is well-known: $\gamma_{NS}^{(0)}(n)=
(8/3)[4\sum_{j=1}^{n}(1/j)-2/n(n+1)-3]$.
In the cases of both $F_2$ and $xF_3$ SFs the numerical expressions
for $\gamma_{NS}^{(1)}(n)$-coefficients are given in Table 1.

\begin{center}
\begin{tabular}{||r|c|c|c|c|c|}
\hline
n  & $\gamma_{NS,F_2}^{(1)}(n)$&$\gamma_{NS,F_3}^{(1)}(n)$
& $\gamma_{NS}^{(2)}(n)$& $\gamma_{NS}^{(3)}(n)|_{[1/1]}$
& $\gamma_{NS}^{(3)}(n)|_{[0/2]}$ \\
\hline
          1&    2.5575 &       0     &           306.6810 & ? & ?    \\
          2&   71.3744&      71.2410 &           612.0598 & 5259 & 5114 \\
          3&  100.8013&      100.7819 &          837.4264 & 6959 & 6900  \\
          4&  120.1446&      120.1401 &         1005.8235 & 8421 & 8414  \\
          5&  134.9049&      134.9035 &         1135.8235 & 9563 & 9562  \\
          6&  147.0028&      147.0023 &         1242.0056 & 10493 & 10482 \\
          7&  157.3323&      157.3321 &         1334.0017 & 11310 & 11280  \\
          8&  166.3862&      166.3861 &         1417.4506 & 12077 & 12012 \\
          9&  174.4683&      174.4682 &         1493.5205 & 12784 & 12706  \\
         10&  181.7808&      181.7808 &         1559.0048 & 13370 & 13271 \\
         11&  188.4662&      188.4662 &          ?  & ? & ? \\
         12&  194.6293&      194.6293 &         ?   & ? & ? \\
         13&  200.3496&      200.3496 &         ?   & ? & ? \\
         14&  205.6891&      205.6891 &         ?   & ? & ? \\
\hline
\end{tabular}

\end{center}
{{\bf Table 1.} The used numerical expressions for the NLO
and NNLO coefficients of the anomalous dimensions of the moments of
the NS SFs at $f=4$ number of flavours and the N$^3$LO Pad\'e estimates.}
\vspace{0.5cm}

These results are normalized to the world with
$f=4$ numbers of active flavours. In the same Table the numerical
expressions for $\gamma_{NS}^{(2)}(n)$, used in the process of the fits,
are presented. In the cases of $n=2,4,6,8,10$ they are following from the
explicit calculations of $\gamma_{NS,F_2}^{(2)}(n)$-terms \cite{LRV1},
normalized to $f=4$, while the $n=3,5,7,9$ numbers were modeled using
the smooth interpolation procedure, originally proposed in Ref.\cite{PKK}.
Note in advance, that since $\gamma_{NS,F_3}^{(2)}(n)$-coefficients
differ from $\gamma_{NS,F_2}^{(2)}(n)$-terms, though by the small
additional contributions (for discussions see Ref.\cite{KKPS1}),
it would be interesting to verify the precision of the  model
for $\gamma_{NS}^{(2)}(n)$, used in the process of our  NNLO
$xF_3$ fits, by the explicit analytical calculations.

Let us now describe the procedure of fixing other theoretical
uncertainties.
After the work of Ref.\cite{SEK} it became rather popular to
model the effects of the higher order terms of perturbative
series in QCD using the expanded Pad\'e approximants.

In the framework of this
technique the values of the terms $C^{(3)}(n)$  and $r(n)$  could be
expressed as

\begin{eqnarray}
Pade~[1/1]~: && C^{(3)}(n)=[C^{(2)}(n)]^2/C^{(1)}(n)        \\
&&   r(n)=q(n)^2/p(n) \\
Pade~[0/2]~: &&
C^{(3)}(n)=2C^{(1)}(n)C^{(2)}(n)     -[C^{(1)}(n)]^3
\\
&&   r(n)= 2p(n)q(n)-[p(n)]^3
\end{eqnarray}

The numerical expressions for $p(n)$ and $q(n)$, obtained from the
results of Table 1 and definitions of Eqs.(9)-(11), together
with the values of the coefficients $C^{(1)}(n)$ and $C^{(2)}(n)$
(which are coming from the calculations of Ref.\cite{WZ}), are presented
in Table 2.

\begin{center}
\begin{tabular}{||r|c|c|c|c|c|c|c|c||} \hline
n&
$p(n)$ & $q(n)$ & $r(n)|_{[1/1]}$ & $r(n)|_{[0/2]}$ &
$C^{(1)}(n)$ & $C^{(2)}(n)$ & $C^{(3)}(n)|_{[1/1]}$ &
$C^{(3)}(n)|_{[0/2]}$ \\ \hline
2& 1.646 & 4.232  & 10.829 &  9.476 & -1.778 & -47.472 & -1267.643 & 174.408
     \\
3& 1.941 & 4.774  & 11.738    & 11.218 & 1.667 & -12.715 & 97.004 & -47.013
          \\
4& 2.050  & 5.546      & 15.003  & 14.123 & 4.867 & 37.117 & 283.085& 246.009
 \\
5&  2.115 &   6.134      & 17.790 & 16.486 & 7.748 & 95.408 & 1174.834 &
1013.328                    \\
6 & 2.165 &  6.595      & 20.087 & 18.407 & 10.351 & 158.291 & 2420.569 &
2167.903
          \\
7&  2.210 &   7.039     & 22.421 & 20.318 & 12.722 & 223.898 &
3940.284 & 3637.790                     \\
8&  2.252 &  7.525 & 25.138 & 22.471 & 14.900 & 290.884 & 5678.657 &
5360.371
                    \\
9& 2.294 &  8.018     &  28.027 &  24.715 & 16.915 & 358.587 &
7601.720 & 7291.305
         \\
10 & 2.334 &  8.375     &  30.049 & 26.382 & 18.791 & 426.442 & 9677.390
&         9391.308                     \\
\hline
\end{tabular}
\end{center}

{{\bf Table 2.} The  expressions for the NLO and NNLO
QCD contributions, used in our fits, and the N$^3$LO Pad\'e estimates.}
\vspace{0.5cm}

In the same Table we also give the estimates for $r(n)$
and $C^{(3)}(n)$, obtained using the expanded [1/1] and [0/2]
Pad\'e approximants formulae of Eqs.(16)-(19).
In the last two columns of Table 1 the estimates for the
N$^3$LO contributions to the anomalous dimension function
$\gamma_{NS}^{(n)}(A_s)$, obtained with the help of the expanded [1/1] and
[0/2] Pad\'e approximants are presented.
One can see that the results of applications of [1/1] and [0/2]
Pad\'e approximants for $\gamma_{NS}^{(3)}(n)$ are almost identical
to each other.

Using the numbers, presented in Table 1, one can construct Pad\'e
motivated expressions for $r(n)$  by substituting the estimates
for $\gamma_{NS}^{(3)}(n)$ into  Eq.(11). It should be stressed, that
the obtained by this way estimates for $r(n)$ will qualitatively agree with
the ones, presented in Table 2 within the ``Pad\'e world'' only, namely
only in the case of application in Eq.(11) of the [1/1] or [0/2] Pad\'e
estimate for the four-loop coefficient of the QCD $\beta$-function $\beta_3$.
However, in the case of $f=4$ the direct application of the [1/1] and [0/2]
Pad\'e approximants underestimates the
calculated value of $\beta_3$ by over the factor
2.5 $(\beta_3|_{[1/1]}=3216.66...;\beta_3|_{[0/2]}=3058.38..).$
In view of this the application of Eq.(11) with the Pad\'e estimated values
of $\gamma_{NS}^{(3)}(n)$ and the explicit expression for $\beta_3$-coefficient
are giving estimates of $r(n)$, drastically different from the ones
presented in Table 2 (for example, for the case of application
of [0/2] P\'ade estimates it gives $r(2)\approx 18.83$,....,$r(10)\approx
55.70$).
It is already known that the accuracy of the  estimates of the
N$^3$LO coefficient of the QCD $\beta$-function can be improved
by some additional fits of polynomial dependence of $\beta_3$ on the number
of flavours $f$ and applying the asymptotic Pad\'e approximant
(APAP) formula
\cite{EKSbet}. Therefore, it might be interesting to think about
the possibility of putting bold
guess Pad\'e estimates of N$^3$LO contributions
to $\gamma_{NS}(A_s)$ (see Table 1) on more solid background.
The analogous steps were already done
in Ref.\cite{Padad}
in the case of the analysis of the status of N$^3$LO Pad\'e
estimates for the anomalous dimension
function of quark mass
and the agreement with the  explicitly  calculated
at this level of QCD the results of Ref.\cite{gamam}  turned
out to be reasonable. One can hope, that the application
of the similar procedure for the APAP estimates of
$\gamma_{NS}^{(3)}$-terms and the substitution of the results
obtained into Eq.(11) (together with the explicit expression
for the $\beta_3$-term) might improve the agreement with the results,
presented in Table 2, which  at this stage we consider as the most
stable results for the N$^3$LO fits.

It should be also stressed that the uncertainties of the
values of $r(n)$ are not so important, since the results of our fits
will be more sensitive to the form of the
Pad\'e approximations predictions for the
contributions into the
coefficient function (namely $C^{(3)}(n)$-terms).

From the results presented in Table 2 one can conclude,
that the theoretical series for $C_{NS}^{(n)}$ for large $n$ $(n\geq 4)$,
which corresponds to the behavior of $xF_3(x,Q^2)$ SF in the intermediate
and large $x$-region, probably have sign
constant structure with asymptotically increasing positive coefficients.
Therefore, the applications of the expanded [1/1] and [0/2] Pad\'e
approximants are
giving theoretical estimates for the terms $C^{(3)}(n)$ $(n\geq 4)$,
which in both cases
have the same positive sign and the same order of magnitude.

However,
in the cases of $n=2,3$ our intuition does not give us the idea what
might be the sign and order of magnitude of the third term in perturbative
series  $C_{NS}^{(2)}(A_s)=1-1.78A_s-47.47A_s^2$ and
$C_{NS}^{(3)}(A_s)=1+1.67A_s-12.71A_s^2$.
Indeed, in these two cases the manipulations with [1/1] and [0/2]
Pad\'e approximants are giving drastically different estimates
for the terms $C^{(3)}(n)$, which in the cases of $n=2,3$ differ
both by sign and value (see Table 2).
These facts, and the structure of the NNLO perturbative series
for $C_{NS}^{(3)}(A_s)$ especially, might indicate that
this series is not yet in the asymptotic regime. Another possibility
is that its coefficients do not have the $(+1)^{n} n!$ growth, but poses
some zigzag structure, which is manifesting itself in the cases of definite
perturbative series of quantum field theory models (for discussions see
e.g. Ref.\cite{BL}). This might give the additional theoretical
uncertainties of modeling higher-order perturbative QCD predictions
for $F_3(x,Q^2)$ in the region of small $x$.

In view of the questionable asymptotic behavior of the NNLO
series for the coefficient functions of NS moments with low $n$
($n=2,3$) we are also
using the  idea of Ref.\cite{Sid} and
consider the results of applications of
non-expanded Pad\'e approximants in the process of the
analysis of the DIS data.

Let us remind that
the corresponding non-expanded [1/1] Pad\'e approximants can be defined as
\begin{equation}
AD(n)|_{[1/1]}=\frac{1+a_1^{(n)}A_s}{1+b_1^{(n)}A_s}
\end{equation}
\begin{equation}
C_{NS}^{(n)}(A_s)|_{[1/1]}=\frac{1+c_1^{(n)}A_s}{1+d_1^{(n)}A_s}
\end{equation}
where $a_1^{(n)}=\bigg([p(n)]^2-q(n)\bigg)/p(n)$, $b_1^{(n)}=-q(n)/p(n)$
and $C_1^{(n)}=\bigg([C^{(1)}(n)]^2-C^{(2)}(n)\bigg)/C^{(1)}(n)$,
$d_1^{(n)}=-C^{(2)}(n)/C^{(1)}(n)$.

The explicit expressions
for the non-expanded [0/2] Pad\'e approximants read:
\begin{equation}
AD(n)|_{[0/2]}=\frac{1}{1+b_1^{(n)}A_s+b_2^{(n)}A_s^2}
\end{equation}
\begin{equation}
C_{NS}^{(n)}(A_s)|_{[0/2]}=\frac{1}{1+d_1^{(n)}A_s+d_2^{(n)}A_s^2}
\end{equation}
where $b_1^{(n)}=-p(n)$, $b_2^{(n)}=p(n)^2-q(n)$, $d_1^{(n)}=-C^{(1)}(n)$
and $d_2^{(n)}=[C^{(1)}(n)]^2-C^{(2)}(n)$.
Since we consider the applications of both [1/1] and [0/2] Pad\'e
approximants as the attempts to model the behavior of
the perturbative
series for the NS Mellin moments beyond the NNLO level, we will use
in Eqs.(20)-(23)  the N$^3$LO expression for the coupling constant $A_s$,
defined  through Eqs.(12)-(14). It is worth to mention here, that
quite recently   the expanded and non-expanded
Pad\'e approximants
were successfully used for the study of     the N$^3$LO approximation
of the ground state energy in quantum mechanics \cite{PP} and
of the behavior of the $\beta$-function
for the quartic Higgs coupling in the Standard Electroweak Model
\cite{Durand}.

The next step is the reconstruction of the structure
function $xF_3(x,Q^2)$ with taking into account
both target mass corrections and twist-4 terms.
The reconstructed SF can be expressed as:
\begin{eqnarray}
xF_{3}^{N_{max}}(x,Q^2)&=&
x^{\alpha}(1-x)^{\beta}
\sum_{n=0}^{N_{max}}
\Theta_n ^{\alpha , \beta}
(x)\sum_{j=0}^{n}c_{j}^{(n)}{(\alpha ,\beta )}
M_{j+2,xF_3}       \left ( Q^{2}\right ) \\ \nonumber
&&+\frac{h(x)}{Q^2}
\label{Jacobi}
\end{eqnarray}
where $\Theta_n^{\alpha,\beta}$ are the Jacobi polynomials and
$\alpha,\beta$ are their parameters, fixed by the condition of the
requirement of the minimization of the error of the reconstruction of
the SF.
In order to take into account the target mass corrections the
Nachtamnn moments
\begin{eqnarray}
M_{n,xF_3}\rightarrow
M_{n,xF_3}^{TMC}(Q^2)=\int_{0}^{1}\frac{dx\xi^{n+1}}{x^2}F_3(x,Q^2)
\frac{1+(n+1)V}{(n+2)},
\label{f3}
\end{eqnarray}
can be used,
where
$\xi=2x/(1+V)$, $V=\sqrt{1+4M_{nucl}^2x^2/Q^2}$ and
$M_{nucl}$ is the mass of a nucleon.
However, to simplify the analysis it is convenient
to expand
equation  (\ref{f3})  into a series in
powers of $M_{nucl}^2/Q^2$ \cite{GEORGI}. Taking into account the
order $O(M_{nucl}^4/Q^4)$ corrections, we get

\begin{eqnarray}
M_{n,xF_3}^{TMC}(Q^2)&=&M_{n,xF_3}^{NS}(Q^2)+\frac{n(n+1)}{n+2}
\frac{M_{nucl.}^2}{Q^2}
M_{n+2,xF_3}^{NS}(Q^2) \\ \nonumber
&&+\frac{(n+2)(n+1)n}{2(n+4)}
\frac{M_{nucl.}^4}{Q^4}M_{n+4,xF_3}^{NS}(Q^2)
+O(\frac{M_{nucl}^6}{Q^6}),
\label{m3}
\end{eqnarray}

We have checked that  the influence  of the order
$O(M_{nucl}^4/Q^4)$ terms in Eq.(26)
to the outcomes  of the concrete fits is very small. Therefore, in what
follows we will use only the first two terms in the r.h.s. of
Eq.(26).

The form of the twist-4 contributions $h(x)$ in Eq.(19) was
first fixed as
\begin{equation}
h(x)=x^{\alpha}(1-x)^{\beta}\sum_{n=0}^{N_{max}} \Theta_n^{\alpha,\beta}(x)
\sum_{j=0}^{(n)}c_j^{(n)}(\alpha,\beta)M_{j+2,xF_3}^{IRR}(Q^2)
\end{equation}
where $c_j^{(n)}(\alpha,\beta)$ are the polynoms, which contain
$\alpha$ and $\beta$-dependent Euler $\Gamma$-functions
and
\begin{equation}
M_{n,xF_3}^{IRR}(Q^2)=\tilde{C}(n)M_{n,xF_3}^{NS}(Q^2)A_2^{'}+
O(\frac{1}{Q^2})
\end{equation}
with $A_2^{'}$ taken as the free parameter and $\tilde{C}(n)$ defined
following the IRR model estimates of Ref.\cite{DW}
as $\tilde{C}(n)=-n-4+2/(n+1)+4/(n+2)+4S_1(n)$ $(S_1(n)=\sum_{j=1}^{n}
1/j)$.
It should be stressed that the appearance  of the
multiplicative
QCD expression  $M_{n,xF_3}^{NS}(Q^2)$ in Eq.(28), generally speaking
different
from the intrinsic coefficients function of the twist-4
contribution,
is leading to definite theoretical
uncertainties
in  the contributions of higher-order QCD
corrections to the twist-4 part of $xF_3(x,Q^2)$.
This could provide the  additional  theoretical errors
in the studies of the the status of the IRR-model predictions for the
twist-4
terms at the NNLO and beyond.

In order to study this question at more definite theoretical level
it is instructive to consider the function
$h(x)$  as the free parameters of the fits, not related to
IRR-model estimates.

We will  estimate the
uncertainties of the values of $\Lambda_{\overline{MS}}^{(4)}$,
 $\alpha_s(M_Z)$ IRR-model parameter $A_2^{'}$ and the twist-4 function
$h(x)$ due to the inclusion of the definite explicitly
calculated N$^3$LO QCD
corrections in the fits and modeling other ones with the help of
the Pad\'e resummation technique.
Our aim will be also the verification of the stability
of the results of Ref.\cite{KKPS2} and the study of the possible
reason of the peculiar behavior of $h(x)$ for $xF_3$ SF, discovered
in Ref.\cite{KKPS2} at the NNLO level.

{\bf 3 (a). The results of the analysis of the experimental data:
the extraction of $\Lambda_{\overline{MS}}^{(4)}$ vs $\alpha_s$ value.}

The definite results for our NLO and NNLO fits,
made for the case of $f=4$ number of active flavours, are
presented in Table 1 of Ref.\cite{KKPS1}, where  the
values of the parton distribution parameters $A,b,c,\gamma\neq$0
are also given. In the present Section we study
the influence of the Pad\'e-motivated estimates of the
N$^3$LO expressions for anomalous dimensions and corresponding
coefficient function
(written down both in the expanded and non-expanded forms) to the results
of  the fits  , which are
resulting in the extractions of the parameter
$\Lambda_{\overline{MS}}^{(4)}$ (and thus $\alpha_(M_Z)$), and of the
common factor $A_2^{'}$ of the IRR model.

It should be stressed that despite the general theoretical preference
of applications of the diagonal Pad\'e approximants
(for the recent analysis see e.g. Ref.\cite{diagonal})
the N$^3$LO [1/1] Pad\'e
approximant description of the CCFR'97 experimental data turned out
to be not acceptable in our case, since  it produces rather high value
of $\chi^2$: $\chi^2/{nep}>2$ (where $nep=86$ is the number
of the experimental points, taken into account in the case of the cut
$Q^2>5~GeV^2$).
However, the application of [0/2] Pad\'e approximants produced
reasonable results.
We think that the nonapplicability of the [1/1]
Pad\'e method in the process of fitting CCFR $xF_3$ data using the
Jacobi polynomial approach can be related to the manifestation of rather
large value of the ratio $[C^{(3)}(2)/C^{(2)}(2)]|_{[1/1]}$
in the expression
for NS moment $M_{2,xF_3}^{NS}$.

The similar
effect of the preference of the [0/2] Pad\'e approximant analysis
over the [1/1] one was found in Ref.\cite{EGKS} in the case
of the comparison
of the QCD
theoretical predictions for the polarized Bjorken sum rule (which
are closely related to the QCD predictions for the first
moment of the $xF_3$ SF, namely
for the Gross-Llewellyn Smith sum rule)
with the available experimental data.

The results for $\Lambda^{(4)}_{\overline{MS}}$, obtained at the LO,
NLO, NNLO and N$^3$LO, modeled by the expanded and non-expanded
Pad\'e approximants,
are presented in Table 3 in the cases of both
$\gamma\neq$0 and $\gamma=0$.

\begin{table}
%
%
\begin{tabular}{||c||c|c|c|c|c|c||} \hline
\hline
 &\multicolumn{3}{c|}{ $\gamma$ - free  } &\multicolumn{3}{c||}{ $\gamma = 0
$ - fixed  }  \\
\hline\hline
                $ Q^2 > $           &
                $\Lambda_{\overline{MS}}^{(4)}$ (MeV)     &
                $ A_2^\prime$(HT)   &
                $ \chi^2$/points    &
                $\Lambda_{\overline{MS}}^{(4)}$ (MeV)     &
                $ A_2^\prime$(HT) ($GeV^2)$)   &
                $ \chi^2$/points    \\
\hline\hline
 5 $GeV^2$ & &                 & & & &  \\
LO & 266$\pm$37      & -- & 113.2/86            &  231$\pm$37      & -- &
126.8/86              \\
   & 436$\pm$56      & -0.33$\pm$0.06 & 82.8/86 &  367$\pm$59      &
-0.30$\pm$0.06 & 102.0/86  \\
NLO & 341$\pm$41     & -- & 87.1/86             &  278$\pm$33      & -- &
109.3/86              \\
   & 371$\pm$31      & -0.12$\pm$0.05 & 81.8/86 &  301$\pm$36      &
-0.10$\pm$0.06 & 105.7/86  \\
NNLO & 293$\pm$29    & -- & 78.4/86             &  258$\pm$28      & -- &
100.6/86              \\
   & 293$\pm$29      & -0.01$\pm$0.05 & 78.4/86 &  258$\pm$29      &
-0.01$\pm$0.05 & 100.5/86  \\
N$^3$LO & 305$\pm$32 & -- & 79.0/86      &  306$\pm$33  & -- &
88.5/86     \\
(n.e.)      & 308$\pm$32 &  -0.03$\pm$0.05  & 78.5/86 & 310$\pm$33 &
 -0.04$\pm$0.05 & 87.8/86  \\
N$^3$LO & 293$\pm$28 & -- & 79.0/86             &  295$\pm$30      & -- &
87.1/86               \\
   & 294$\pm$28      & -0.02$\pm$0.05 & 78.8/86 &  297$\pm$29      &
 -0.03$\pm$0.05 & 86.8/86    \\
\hline
 10 $GeV^2$ & &                 & & & &  \\
LO & 287$\pm$37      & -- & 77.7/63             &  279$\pm$40      & -- &
79.6/63               \\
   & 531$\pm$74      & -0.52$\pm$0.11 & 58.0/63 &  503$\pm$97      &
-0.49$\pm$0.15 & 61.7/63   \\
NLO & 350$\pm$36     & -- & 64.4/63             &  328$\pm$39      & -- &
69.1/63               \\
   & 439$\pm$55      & -0.24$\pm$0.10 & 58.6/63 &  392$\pm$173     &
-0.19$\pm$0.27 & 65.3/63   \\
NNLO & 308$\pm$34    & -- & 58.7/63             &  285$\pm$34      & -- &
68.1/63               \\
  & 313$\pm$37       & -0.03$\pm$0.09 & 58.6/63 &  285$\pm$38      &
-0.01$\pm$0.09 & 68.1/63   \\
N$^3$LO & 303$\pm$33  & -- & 60.4/63             &  303$\pm$33   & -- &
63.0/63     \\
(n.e.)        & 309$\pm$36  & -0.03$\pm$0.09 & 60.3/63 & 310$\pm$37    &
-0.04$\pm$0.09 & 62.8/63  \\
N$^3$LO & 295$\pm$30 & -- & 59.9/63             &  296$\pm$31      & -- &
62.2/63               \\
     & 297$\pm$33    & -0.01$\pm$0.09 & 59.8/63 &  298$\pm$34      &
-0.02$\pm$0.09 & 62.2/63   \\
\hline
 15 $GeV^2$ & &                 & & & &  \\
LO & 319$\pm$41      & -- & 58.8/50             &  318$\pm$41      & -- &
58.9/50               \\
   & 531$\pm$57      & -0.57$\pm$0.13 & 50.2/50 &  517$\pm$112     &
-0.54$\pm$0.27 & 50.9/50   \\
NLO & 365$\pm$40     & -- & 53.0/50             &  355$\pm$44      & -- &
54.1/50               \\
   & 441$\pm$44      & -0.25$\pm$0.13 & 50.9/50 &  408$\pm$43      &
-0.19$\pm$0.13 & 52.8/50   \\
NNLO & 314$\pm$37    & -- & 50.9/50             &  294$\pm$36      & -- &
56.7/50               \\
   & 308$\pm$45      &  0.03$\pm$0.14 & 50.8/50 &  279$\pm$44      &
0.09$\pm$0.15 & 56.3/50   \\
N$^3$LO & 304$\pm$36  & -- & 52.8/50  & 303$\pm$36 & -- & 54.3/50 \\
(n.e.)        & 297$\pm$43 & 0.05$\pm$0.14 & 52.7/50 & 294$\pm$43 & 0.05$\pm$0.14
    & 54.1/50 \\
N$^3$LO & 296$\pm$33 &                & 52.3/50 &  295$\pm$33      &      --
& 53.8/50   \\
   & 286$\pm$38      &  0.07$\pm$0.14 & 52.0/50 &  283$\pm$39      &
0.08$\pm$0.14 & 53.4/50   \\
\hline \hline
\end{tabular}
{{\bf Table 3}. The results of the extractions
of the parameter $\Lambda_{\overline{MS}}^{(4)}$ and
the IRR coefficient $A_2^{'}$, (in $GeV^2$) defined in Eq.(23), from LO, NLO, NNLO
and N$^3$LO
non-expanded (n.e.) and expanded Pad\'e
fits of CCFR'97 data.}
\end{table}
\vspace{0.3cm}

Looking carefully on Table 3 we arrive to the following
conclusions:

\begin{itemize}

\item Our fits demonstrate that
the effects of the NNLO perturbative QCD contributions
are   important in the analysis of the CCFR data.
Indeed, for different $Q^2$-cuts they are diminishing the
values of the QCD scale parameter $\Lambda_{\overline{MS}}^{(4)}$
by the contribution, which is varying in the range over
$50-120~MeV$, provided twist-4 corrections are taken into account
through the IRR model of Ref.\cite{DW};

\item The values of $\Lambda_{\overline{MS}}^{(4)}$, which are coming
from the fits with taking into account the [0/2] Pad\'e
estimates (both in the expanded and non-expanded  variants)
turn out to be really nonsensitive to choosing the $Q^2$-cut
of the data, fixation of the value of $\gamma$ and thus incorporation
of $(1+\gamma x)$-factor in the parton distribution model.
This
in its turn can indicate that the change of
the used by us model
$xF_3(x,Q_0^2)=A(Q_0^2)x^{b(Q_0^2)}(1-x)^{c(Q_0^2)}
(1+\gamma(Q_0^2))x)$ to $xF_3(x,Q_0^2)=A(Q_0^2)x^{b(Q_0^2)}
(1+\gamma(Q_0^2)x)+\epsilon(Q_0^2)\sqrt{x})$, used in the
MRST and GRV fits, should not affect significantly the results obtained;

\item The large errors in the definite LO and NLO results for
$\Lambda_{\overline{MS}}^{(4)}$, presented in Table 3,
are reflecting the correlations of these values with the errors of
parton distributions parameters, which will be not considered in this
paper and presented by us elsewhere ;

\item
For all $Q^2$-cuts and values of the parameter $\gamma$ the applications
of the N$^3$LO fits performed with the help of the expanded
 [0/2] Pad\'e approximants technique result in the smaller value of
$\Lambda_{\overline{MS}}^{(4)}$ and slightly smaller $\chi^2$ value than in the
case of application of the non-expanded [0/2] Pad\'e approximants.
In our future studies we will consider the results of application
of both expanded and non-expanded Pad\'e approximations.

\item
For all $Q^2$-cuts the expanded N$^3$LO results for
$\Lambda_{\overline{MS}}^{(4)}$ is smaller then  the similar results
for non-expanded Pad\'e fits by  10-15 $MeV$. This effect
is coming from the expansion of Pad\'e approximants in Taylor series.
The  difference in the results of the expanded and non-expanded Pad\'e
approximants can be considered as the estimate of the part of theoretical
error of the N$^3$LO results.
\end{itemize}

Using the LO, NLO, NNLO and N$^3$LO variants of the rigorous
$\overline{MS}$-scheme matching conditions, derived in Ref.\cite{CKS}
following the lines of Ref.\cite{BW}, we  transform
$\Lambda_{\overline{MS}}^{(4)}$ values
through the threshold of the production of the fifth flavour
$M_5=m_b$ (where $m_b$ is the $b$-quark pole mass) and  obtain
the related values of $\Lambda^{(5)}_{\overline{MS}}$
with the help of  the following equation:
\begin{eqnarray}
\beta_0^{f+1}ln\frac{\Lambda
_{\overline{MS}}^{(f+1)~2}}{\Lambda_{\overline{MS}}^{(f)~2}}=
(\beta_0^{f+1}-\beta_0^f)L_h \\ \nonumber
+\delta_{NLO}+\delta_{NNLO}+
\delta_{N^3LO}
\end{eqnarray}
\begin{equation}
\delta_{NLO}=\bigg(\frac{\beta_1^{f+1}}{\beta_0^{f+1}}-\frac{\beta_1^f}
{\beta_0^f}\bigg)ln L_h-\frac{\beta_1^{f+1}}{\beta_0^{f+1}}ln
\frac{\beta_0^{f+1}}{\beta_0^f}
\end{equation}
\begin{eqnarray}
\delta_{NNLO}=\frac{1}{\beta_0^f L_h}\bigg[\frac{\beta_1^f}{\beta_0^f}
\bigg(\frac{\beta_1^{f+1}}{\beta_0^{f+1}}
-\frac{\beta_1^f}{\beta_0^f}\bigg)ln L_h
\\ \nonumber
+\bigg(\frac{\beta_1^{f+1}}{\beta_0^{f+1}}\bigg)^2
-\bigg(\frac{\beta_1^{f}}{\beta_0^{f}}\bigg)^2
-\frac{\beta_2^{f+1}}{\beta_0^{f+1}}
+\frac{\beta_2^{f}}{\beta_0^{f}}-C_2\bigg]
\end{eqnarray}
\newpage
\begin{eqnarray}
\delta_{N^3LO}=\frac{1}{(\beta_0^f L_h)^2}\Bigg[
-\frac{1}{2}\bigg(\frac{\beta_1^f}{\beta_0^f}\bigg)^2
\bigg(\frac{\beta_1^{f+1}}{\beta_0^{f+1}}-\frac{\beta_1^f}{\beta_0^f}\bigg)
ln^2 L_h
\\ \nonumber
+\frac{\beta_1^f}{\beta_0^f}\bigg[-\frac{\beta_1^{f+1}}{\beta_0^{f+1}}
\bigg(\frac{\beta_1^{f+1}}{\beta_0^{f+1}}-\frac{\beta_1^f}{\beta_0^f}\bigg)
+\frac{\beta_2^{f+1}}{\beta_0^{f+1}}
-\frac{\beta_2^f}{\beta_0^f}+C_2\bigg]ln L_h
\\ \nonumber
+\frac{1}{2}\bigg(-\bigg(\frac{\beta_1^{f+1}}{\beta_0^{f+1}}\bigg)^3
-\bigg(\frac{\beta_1^f}{\beta_0^f}\bigg)^3-\frac{\beta_3^{f+1}}{\beta_0^{f+1}}
+\frac{\beta_3^f}{\beta_0^f}\bigg)
\\ \nonumber
+\frac{\beta_1^{f+1}}{\beta_0^{f+1}}\bigg(\bigg(
\frac{\beta_1^f}{\beta_0^f}\bigg)^2+
\frac{\beta_2^{f+1}}{\beta_0^{f+1}}-\frac{\beta_2^f}{\beta_0^f}+C_2\bigg)
-C_3\Bigg]
\end{eqnarray}
where $C_2=-7/24$ was calculated in Ref.\cite{LRVS} and the analytic
expression for $C_3$, namely  $C_3=-(80507/27648)\zeta(3)-(2/3)\zeta(2)
((1/3) ln2+1)-58933/124416 +(f/9)[\zeta(2)+2479/3456]$ was recently found
in Ref.\cite{CKS}. Here
$\beta_i^f$ ($\beta_i^{f+1}$) are the coefficients of the
$\beta$-function with $f$ ($f+1$)
numbers of active flavours,
$L_h=ln(M_{f+1}^2/\Lambda_{\overline{MS}}^{(f)~2})$ and $M_{f+1}$ is the
threshold of the production of the quark of $(f+1)$-th flavour.
In our analysis we should take $f=4$ and $m_b\approx 4.6~GeV$.

In the case of the non-zero values of the twist-4 function
$h(x)\neq 0$ the results of the LO, NLO, NNLO and N$^3$LO
fits, made both in the expanded and non-expanded versions
of the [0/2] Pad\'e motivated approach are presented in Table 4.

It should be stressed that we are considering the outcomes of our N$^3$LO
approximated fits as the rate of theoretical uncertainties  of the
NNLO results in the same manner like the results of the NNLO analysis
will be considered as the measure of theoretical uncertainties
of the NLO results.
In particular, we will introduce the characteristic deviations
$\Delta^{NNLO}$=
$|(\Lambda_{\overline{MS}}^{(4)})^{N^3LO}-
(\Lambda_{\overline{MS}}^{(4)})^{NNLO}|$,
$\Delta^{NLO}$=
$|(\Lambda_{\overline{MS}}^{(4)})^{NNLO}-
(\Lambda_{\overline{MS}}^{(4)})^{NLO}|$.

It is worth to emphasize that the results for
the Pad\'e motivated N$^3$LO results for
$\Lambda_{\overline{MS}}^{(4)}$ and thus $\alpha_s(M_Z)$ became
closer to  the NNLO ones (provided the statistical
error bars are taken into account, see Table 1). Moreover, the difference
$\Delta^{NNLO}$=
$|(\Lambda_{\overline{MS}}^{(4)})^{N^3LO}-
(\Lambda_{\overline{MS}}^{(4)})^{NNLO}|$,
is drastically smaller then the NLO correction term
$\Delta^{NLO}$=
$|(\Lambda_{\overline{MS}}^{(4)})^{NNLO}-
(\Lambda_{\overline{MS}}^{(4)})^{NLO}|$. The similar tendency
$\Delta^{NNLO}<<\Delta^{NLO}$ is taking place in the case
of the fits without twist-4 corrections. These observed properties
indicate the reduction of the theoretical errors due to cutting
the analyzed perturbation series at the different orders.

It is known that the inclusion of the higher-order perturbative
QCD corrections into the  comparison with the
experimental data is decreasing the scale-scheme theoretical
errors of the results for $\Lambda_{\overline{MS}}^{(4)}$ and thus
$\alpha_s(M_Z)$ (see e.g. Refs.\cite{ChK,PKK,EGKS}). Among the ways of
probing the scale-scheme uncertainties are the scheme-invariant
methods, namely the principle of minimal sensitivity,
the effective charges approach,
( which is known to be
identical to the  scheme-invariant perturbation theory) and the
BLM approach (for the review of these methods see e.g. Ref.\cite{ChK2}).
The scheme-invariant methods
were already used to estimate the effects of the
unknown higher order corrections in SFs (see Ref.\cite{KPSG}, where
a strong decrease of the value of the QCD scale
parameter was found in the process of the scheme-invariant
fit of the experimental data for $F_2$ SF ) and to try to
predict the unknown
at present N$^3$LO corrections to the definite physical
quantities \cite{KatSt}, and DIS sum rules among others.
Note that the predictions of Ref.\cite{KatSt} are in agreement
with the results of applications of the Pad\'e resummation technique
(see Ref.\cite{SEK}). Therefore, we can conclude that the application
of the methods of the Pad\'e approximants should  lead to the reduction
of the scale-scheme dependence uncertainties of the values of
$\alpha_s(M_Z)$ in the analysis of the CCFR data.

We are presenting now the values of $\alpha_s(M_Z)$, extracted
from the fits of the CCFR'97 experimental data
for the $xF_3$ SF, obtained with fixing  twist-4
contribution through the IRR model of Ref.\cite{DW}:

\begin{eqnarray}
NLO~HT~of \cite{DW}~\alpha_s(M_Z) =0.121 \pm 0.002(stat)
\\ \nonumber
\pm 0.005(syst)
\pm 0.005(theory)
\end{eqnarray}
\begin{eqnarray}
NNLO~HT~of \cite{DW}~\alpha_s(M_Z)=0.118\pm 0.002(stat)
\\ \nonumber
\pm 0.005(syst)
\pm 0.003(theory)~.
\end{eqnarray}

The results of the extractions of $\alpha_s(M_Z)$, with
twist-4 contribution, considered as
the additional free parameters of the fit, have the following
form:

\begin{eqnarray}
NLO~HT~free~\alpha_s(M_Z) =0.124^{+0.007}_{-0.009}(stat)
\\ \nonumber
\pm 0.005(syst)
\pm 0.008(theory)
\end{eqnarray}
\begin{eqnarray}
NNLO~HT~free~\alpha_s(M_Z)=0.116^{+0.006}_{-0.007}(stat)
\\ \nonumber
\pm 0.005(syst)
\pm 0.004(theory)
\end{eqnarray}
where the systematic uncertainties are taken from the
CCFR experimental
analysis, presented in the first work of Ref.\cite{Seligman},
and the theoretical uncertainties in the results
of Eqs.(33),(35) [Eqs.(34),(36)] are estimated by the diffferences
between the
central values of the outcomes of the NNLO and NLO [N$^3$LO and NNLO] fits,
presented in Table 1, plus the arbitrariness in the application of
smoothing procedure of the $\overline{MS}$-scheme
matching condition (which following the
considerations of Ref.\cite{DV} was estimated as $\Delta\alpha_s(M_Z)=
\pm 0.001$) and the uncertainties in $\alpha_s(M_Z)$ due to nuclear effects
on iron target (see discussions in Sec.4). In estimating the theoretical
errors of the inclusion of the N$^3$LO corrections we take into account
the diffferences between the applications of the expanded and non-expanded
Pad\'e approximants.

It can be seen that due to the large overall number of the fitted
parameters the results of Eqs.(35),(36) for
$\alpha_s(M_Z)$ have rather large statistical
uncertainties. As can be seen from the results of Eq.(33),(34)
for the QCD coupling constant
it is possible to decrease their values by fixing the
concrete form of the twits-4 parameter $h(x)$.
However, if one is interested in the fitted form of the twist-4
parameter $h(x)$, one should take for granted these intrinsic
theoretical uncertainties of the value of $\alpha_s(M_Z)$.

{\bf 3 (b). The results of the analysis of the experimental data:
the extraction of parameters of the twist-4 terms.}

Apart of the perturbative QCD contributions the  expressions
for DIS structure functions should contain the power-suppressed
high-twist terms, which reflect the possible non-perturbative
QCD effects. The  studies of these terms have rather
long history. At the beginning of these studies it was
realized that the twist-4 contributions to structure functions
should have the pole-like behavior $\sim 1/((1-x)Q^2)$
\cite{Stw,Nason}. This behavior was used in the
phenomenological
investigations of the earlier less precise than CCFR DIS $\nu N$ data
\cite{Barnett,Isaev,f3jac}, which
together with  different other procedures
on analyzing neutrino DIS data \cite{DeRujula,Yndurain}
was considered as the source of the information about
scaling violation parameters. The development of renormalon
technique (Refs.\cite{IRR,DW} and Ref.\cite{Beneke} for the
detailed review) pushed ahead the more detailed phenomenological
analysis of the possibility of detecting higher-twist components
in the available at present most precise DIS data obtained by
BCDMS, SLAC, CCFR and other collaborations.
It turned out that despite the qualitative status of renormalon
approach the satisfactory description of the results of QCD
NLO $F_2$ SF analysis \cite{VM}
in terms of IRR technique was achieved
\cite{DW,IRR}.
The next step was to clarify the status of the predictions
of Ref.\cite{DW} for the form and sign of the twist-4 contributions
to $xF_3$ SF. The study of this problem was done in
Ref.\cite{KKPS2} (see also Ref.\cite{S}).
In this chapter we are discussing the results of  more refined
analysis of the behavior of the twist-4 contributions  to
$xF_3$ SF at the LO, NLO, NNLO and beyond.

In Table 3 we study the dependence of the extracted value
of the  parameter $A_2^{'}$ from the different orders of
perturbative QCD predictions,
$Q^2$-cuts of the CCFR experimental data and the coefficient $\gamma$
of parton distributions model for $xF_3$. Note, that
the parameter $A_2^{'}$  was introduced in the IRR model
of Eq.(28), taken from Ref.\cite{DW}, and fixed there as $A_2^{'}\approx
-0.2~GeV^2$, which is necessary for the description of the
fitted twist-4 results of Ref.\cite{VM} for $F_2$ within
IRR language. We found that the value of this parameter
extracted at the LO and NLO, is negative, differ from zero for
about one standard deviation and qualitatively in agreement with the
IRR-motivated guess of Ref.\cite{DW}. Moreover, the results of our
LO and NLO fits
are also in agreement with the made in Ref.\cite{f3jac} earlier extraction
from the old DIS neutrino data of
the value  $h=-0.38\pm 0.06~GeV^2$ of the different model of the
twist-4 contribution to $xF_3$, namely $xF_3(x,Q^2)h/((1-x)Q^2)$.

It is interesting to notice that the results of the multiloop extractions
of the parameter $A_2^{'}$ from the CCFR data are almost nonsensitive
to the introduction of the additional parameter $\gamma$ in the parton
distribution model.

Another  observation is  that the results of Table 3
reveal that for larger $Q^2$-cuts $10-15~GeV^2$ the results of
$A_2^{'}$ in the LO and NLO fits are more stable to the inclusion
of
the concrete experimental data, than in the case of the  low
$Q^2$ cut ($5~GeV^2$). This feature can be related to the logarithmic
increase of the QCD coupling constant $A_s$ at lower $Q^2$.
However, since we are interested in the extraction of the
power-suppressed twist-4 contribution, we shall concentrate
on the discussion of the more informative from our point of view
fits with low $Q^2$-cut $5~GeV^2$, which contain more experimental points
and thus are  more statistically motivated.

We  also observed that in the process of multiloop extraction  of
$A_2^{'}$ the tendency $\chi^2_{LO}>\chi^2_{NLO}>\chi^2_{NNLO}
\sim\chi^2_{N^{3}LO}$ takes place.
It should be stressed that in opposite to the
results of the LO and NLO extraction the NNLO value
of $A_2^{'}$ is compatible with zero and is stable to the inclusion
of the N$^3$LO contribution to $A_{s}$ through the Pad\'e approximants
for the NS Mellin moments.

We are now turning to the pure phenomenological
extraction of the twist-4 contribution $h(x)$ to
$xF_3$ (see Eq.(24)), which is motivated by the
work of Ref.\cite{VM} for $F_2$ data. In the framework
of this approach the $x$-shape of $h(x)$ is parametrized by
the additional parameters $h_i=h(x_i)$, where $x_i$ are
the points of the experimental
data bining. The results of the multiloop extractions of these
parameters are presented in Table 4 and are illustrated by
the curves of Fig.1.

\newpage
\begin{table}
\begin{tabular}{|c|c|c|c|c|c|} \hline \hline
                    &      LO            &       NLO       &   NNLO &  N$^3$LO
&  N$^3$LO(n.e.)\\ \hline
 $\chi^2/points$    &  66.2/86           &        65.6/86  &  65.7/86 & 65.6/86
& 64.3/86  \\
  A    &   5.44 $\pm$  1.74  &  3.70 $\pm$  1.56   &  4.54 $\pm$ 0.88 & 5.24 $
\pm$ 0.92 & 5.16 $\pm$ 0.75 \\
  b    &   0.74 $\pm$  0.10  &  0.66 $\pm$  0.11   &  0.69 $\pm$ 0.06 & 0.72 $
\pm$ 0.06 & 0.73$\pm$ 0.05\\
  c    &   4.00 $\pm$  0.18  &  3.78 $\pm$  0.21   &  3.72 $\pm$ 0.19 & 3.58 $
\pm$ 0.26 & 3.48 $\pm$ 0.26\\
$\gamma$  &1.72 $\pm$  1.25  &  2.86 $\pm$  1.72   &  1.43 $\pm$ 0.69 & 0.72 $
\pm$ 0.63 & 0.71 $\pm$ 0.51\\
$\Lambda_{\overline{MS}}^{(4)}$ & 338 $\pm$ 169 & 428 $\pm$ 158  & 264 $\pm$ 85
& 248 $\pm$ 76 & 310 $\pm$ 100\\
 $[MeV]$ &               &                &               &              &
         \\ \hline \hline
   $x_i$                     &\multicolumn{5}{c||}{  $h(x_i)~[GeV^2]$ }   \\
\hline
0.0125 &   0.206 $\pm$ 0.321 &    0.213  $\pm$ 0.332 &     0.170 $\pm$ 0.302 &
0.165  $\pm$ 0.300 & 0.190  $\pm$ 0.303 \\
0.0175 &   0.061 $\pm$ 0.268 &    0.091  $\pm$ 0.289 &     0.030 $\pm$ 0.241 &
0.020  $\pm$ 0.240 & 0.051  $\pm$ 0.247\\
0.025  &   0.146 $\pm$ 0.204 &    0.220  $\pm$ 0.241 &     0.136 $\pm$ 0.178 &
0.118  $\pm$ 0.176 & 0.156  $\pm$ 0.186\\
0.035  &  -0.021 $\pm$ 0.185 &    0.114  $\pm$ 0.240 &     0.008 $\pm$ 0.179 &
-0.021 $\pm$ 0.175 & 0.023  $\pm$ 0.182\\
0.050  &   0.031 $\pm$ 0.142 &    0.245  $\pm$ 0.230 &     0.124 $\pm$ 0.164 &
0.080  $\pm$ 0.154 & 0.130  $\pm$ 0.162\\
0.070  &  -0.145 $\pm$ 0.127 &    0.138  $\pm$ 0.233 &     0.033 $\pm$ 0.165 &
-0.026 $\pm$ 0.146 & 0.028  $\pm$ 0.163\\
0.090  &  -0.177 $\pm$ 0.125 &    0.139  $\pm$ 0.225 &     0.076 $\pm$ 0.166 &
0.008  $\pm$ 0.143 & 0.063  $\pm$ 0.161\\
0.110  &  -0.340 $\pm$ 0.126 &   -0.015  $\pm$ 0.205 &    -0.026 $\pm$ 0.166 &
-0.095 $\pm$ 0.145 & -0.042  $\pm$ 0.163\\
0.140  &  -0.404 $\pm$ 0.114 &   -0.092  $\pm$ 0.147 &    -0.027 $\pm$ 0.141 &
-0.088 $\pm$ 0.128 & -0.048  $\pm$ 0.137\\
0.180  &  -0.350 $\pm$ 0.164 &   -0.077  $\pm$ 0.122 &     0.054 $\pm$ 0.127 &
0.016  $\pm$ 0.130 & 0.021  $\pm$ 0.128\\
0.225  &  -0.554 $\pm$ 0.237 &   -0.348  $\pm$ 0.170 &    -0.167 $\pm$ 0.135 &
-0.174 $\pm$ 0.139 & -0.221  $\pm$ 0.142\\
0.275  &  -0.563 $\pm$ 0.334 &   -0.462  $\pm$ 0.272 &    -0.196 $\pm$ 0.193 &
-0.171 $\pm$ 0.189 & -0.264  $\pm$ 0.206\\
0.350  &  -0.314 $\pm$ 0.418 &   -0.368  $\pm$ 0.371 &     0.070 $\pm$ 0.200 &
0.128  $\pm$ 0.188 & 0.026  $\pm$ 0.206\\
0.450  &  -0.117 $\pm$ 0.415 &   -0.266  $\pm$ 0.401 &     0.183 $\pm$ 0.213 &
0.237  $\pm$ 0.191 & 0.119  $\pm$ 0.209\\
0.550  &   0.087 $\pm$ 0.333 &   -0.109  $\pm$ 0.352 &     0.097 $\pm$ 0.234 &
0.116  $\pm$ 0.224 & -0.051  $\pm$ 0.278\\
0.650  &   0.377 $\pm$ 0.215 &    0.221  $\pm$ 0.244 &     0.259 $\pm$ 0.188 &
0.271  $\pm$ 0.189 & 0.184  $\pm$ 0.220\\
\hline  \hline
\multicolumn{6}{p{15cm}}{{\bf Table 4.}
The results of the LO, NLO
($N_{max}=10$), NNLO and the N$^3$LO ($N_{Max}=6$)
expanded and non-expanded (n.e.) Pad\'e
QCD fit (with TMC)
of the CCFR'97 $xF_3$ SF data
for the values of HT contributions $h(x)$ and for the
parameters $A,b,c,\gamma$ with the corresponding
statistical errors.}\\
\end{tabular}
\end{table}

Looking carefully on Table 4 and Fig.1 we observe the following features:
\begin{enumerate}
\item The $x$-shape of the twist-4 parameter
is not inconsistent with the
expected rise of
$h(x)$ for $x\rightarrow 1$ \cite{Stw,Nason} in all orders of perturbation
theory;
\item The values of the parameters $h(x_i)$ at the upper and lower
points of kinematic region ($x_{16}$=0.650 and $x_{1}$=0.0125)
are stable to the inclusion of the higher order perturbative QCD corrections
and application of the Pad\'e resummation technique. At large values of
$x$ this feature is in agreement with the previous statement;
\item The function  $h(x)$ seems to cross zero twice:
at small $x$ of order 0.03
and larger $x$ about 0.4. It should be noted that the sign-alternating
behavior of the twist-4 contributions to DIS structure functions
was qualitatively predicted in Ref.\cite{Isaev};
\item In the LO and NLO our results are in qualitative agreement
with the IRR prediction of Ref.\cite{DW} (for discussions see
Ref.\cite{Beneke});
\item In the NNLO this agreement is not so obvious, though the certain
tendency of following the general shape of IRR prediction \cite{DW}
still survives;

\newpage

\vspace*{3.0cm}
\epsfxsize=10cm
\epsfysize=8cm
\centerline{\epsfbox{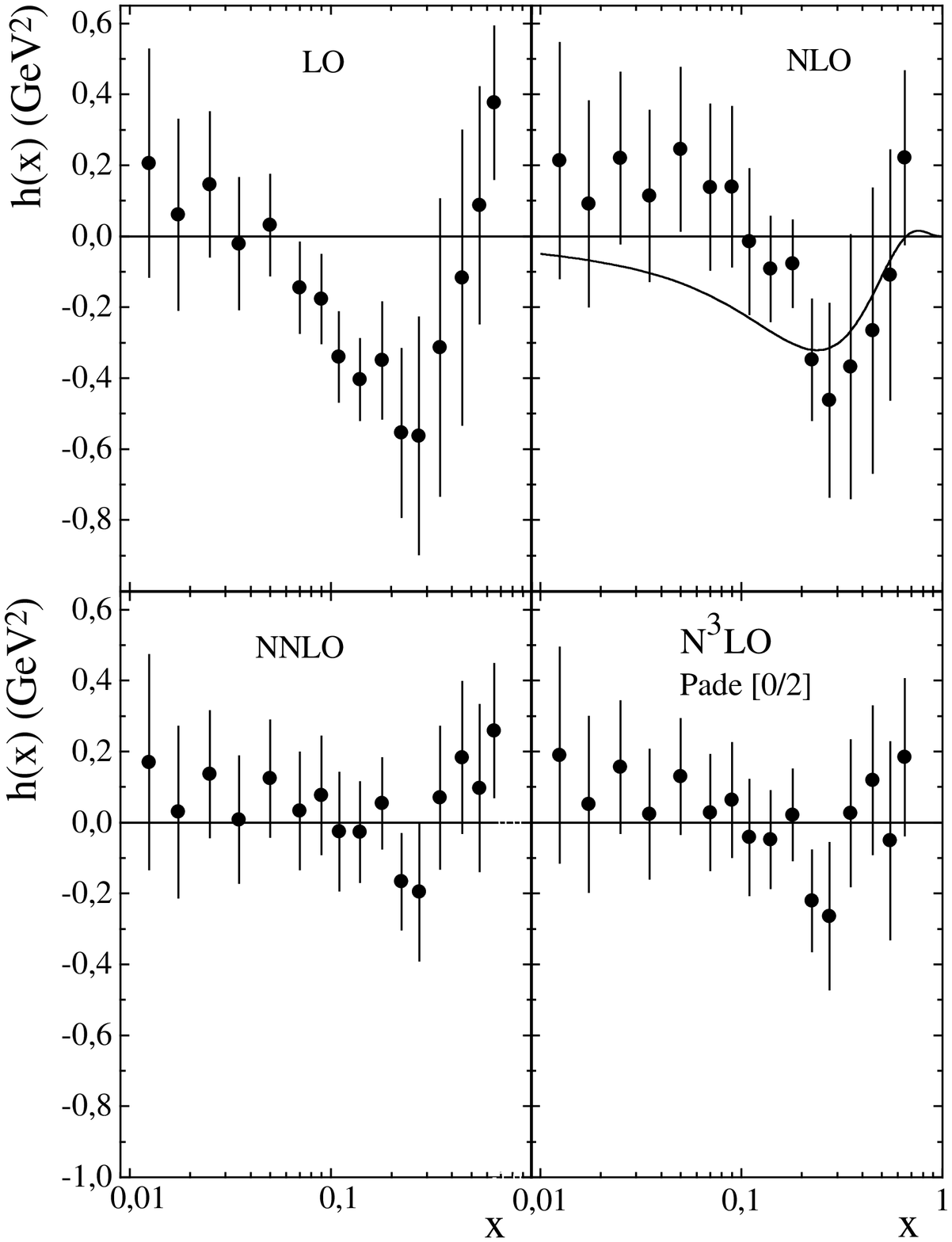}}
  \vspace*{-.5cm}
\begin{center}
Fig.1~The results of the LO, NLO, NNLO and N$^3$LO [0/2]
Pad\'e extarctions of the twist-4 contributions of $h(x)$.
The solid line is  the IRR-model prediction of Ref.\cite{DW}.
\end{center}

\item However, at the NNLO we observe  the partial nullification
of $h(x)$ within statistics error bars. Thus we conclude that the
inclusion of the NNLO corrections into the game is shadowing the
effects of the power suppressed terms,
or that the effects of the twist-4 corrections are
nondetectable at the NNLO.
 This property was previously
observed at the LO as the result of the less precise DIS neutrino
data in Ref.\cite{Barnett}. In the modern experimental situation, namely
in the process of the analysis of the more precise DIS neutrino data
of CCFR collaboration, we are observing this feature at the NNLO;
\item We checked the reliability of the foundation of
partial nullification of $h(x)$ at the NNLO by going beyond this
perturbative approximation using the methods of Pad\'e approximants.
The result of this analysis reveals the stability of the NNLO results
for $h(x)$ and its partial nullification;
\item The property of partial nullification of $h(x)$
at the NNLO and N$^3$LO is identical to the effect
of nullification  of the IRR model parameter $A_2^{'}$
at the NNLO and N$^3$LO (see Table 3);
\item These observed properties clarifies why the results of the NNLO
and N$^3$LO fits for $\Lambda_{\overline{MS}}^{(4)}$ , presented in Tables 3,
practically
do not depend  from the inclusion of the twist-4 contribution
through the IRR model.
Indeed, at this level the twist-4 terms have almost zero effect.
\end{enumerate}

To our point of view  the  foundations (8)-(9)
reflects the selfconsistency of the results of our different
fits with twist-4 included by different ways.

{\bf 4. The quest of  the inclusion of the effects of nuclear
corrections.}

The effects of nuclear corrections
are remaining the important source of the uncertainties of
the analysis of the DIS data.
This  is especially important for the experiments
on heavy targets and in the case of CCFR data--on iron  $^{56}Fe$.

The attempts to study these effects were done in Ref.\cite{SidTok}
in the framework of Deutron-motivated model. The
satisfactory QCD description of the CCFR data for $xF_3$ was achieved
due to the reason that in this case the nuclear effects
do not exceed 5 $\%$ effect. However, the more
realistic description of nuclear effect for $xF_3$
on  $^{56}Fe$ \cite{Kulagin}
revealed the appearance of new $1/Q^2$ and $1/M$ corrections
for NS moments (where $M$ is the mass of the nucleon),
which have the following form
\begin{eqnarray}
M_{n}^{A}(Q^2)/A&=&\bigg(1+\frac{\epsilon}{M}(n-1)+\frac{<\bf{p}^2>}{6M^2}n(n-1)
+ O(\frac{1}{M^3})\bigg)M_n^{NS}(Q^2) \nonumber \\
&&+<\Delta p^2>\partial_{p^2}M_n^{NS}(Q^2) \nonumber \\
&&+\frac{2<\bf{p}^2>}{3Q^2}n(n+1)
M_{n+2}^{NS}(Q^2)
\end{eqnarray}
where for $^{56}Fe$ the parameters of the nuclear model, adopted
in Ref.\cite{Kulagin} are
$<\epsilon>\approx -56 ~MeV$,
$<{\bf p}^2>/(2M)\approx 35~MeV$, $<\Delta p^2>_{Fe}\approx -0.17~GeV$
and the derivative $\partial_{p^2}M_n(Q^2)$ is taking into account that
the target momentum $p$ can be generally of-mass-shell. This effect is
resulting in the following independent from the nuclear content
contribution
\cite{Kulagin}
\begin{equation}
\partial_{p^2}M_n(Q^2)=\partial_{p^2}M_n^{as}+\frac{n}{Q^2}\bigg
(M_n^{NS}+M^2\partial_{p^2}M_n^{as}\bigg)
\end{equation}
where the numerical values of $\partial_{p^2}M_n^{as}$ were also
presented in Ref.\cite{Kulagin}.

Note, that the effects of the nuclear corrections in DIS
were also recently studied in Ref.\cite{Spain1} in the case
of $xF_3$ SF and in Ref.\cite{Spain2},\cite{SL}
in the case of $F_2$ SF (for the earlier related works see e.g.
Ref.\cite{AKV}).
However, in our studies we will concentrate
ourselves on the consideration of the results of Ref.\cite{Kulagin}.

We included the corrections of Eqs.(37)-(38) into our fits and
observed the unacceptable increase of $\chi^2$ value.
We think that this can be related to
to the manifestation of the possible asymptotic character of the
$1/M$-expansion in Eq.(37),
since the third term in the
brackets  of the r.h.s. of Eq.(32) becomes
comparable with the first term (which is equal to  unit)
for the $n \sim 8$ used in our fits. Note that
the moments with large $n$ are important in the
reconstruction of the behavior of the $xF_3$ SF at $x\rightarrow 1$.
This observed feature
necessitates the derivation of the explicit expression
for $M_n^A(Q^2)$, which is not expanded in powers of $1/M$-terms.
It should be noted that the  problem of the possible asymptotic nature
of the power suppressed expansions was mentioned in the case of
Ellis-Jaffe and Bjorken  DIS sum rules  in Ref.\cite{Ioffe}.

Another possibility of the nonconvergence of our fits
with nuclear corrections of Eq.(37) taken into account
might
be related to the fact
that the parton distribution model for the nuclear SF $xF_3^{^{56}Fe}$
can be different from the canonical model, used by us.

In any case  we think that  the problem of the taking into account of the
heavy nuclear
effects in the process of fits of $xF_3$ data is  really on the agenda.

\newpage

{\bf Conclusion}

In this work we presented the results of the extractions
of $\alpha_s(M_Z)$  and twist-4 terms from the QCD analysis
of the CCFR data with taking into account definite QCD
corrections at the NNLO and beyond. Within experimental
and theoretical errors our results for
$\alpha_s(M_Z)$ are in agreement with other extractions of this
fundamental parameter, including its world average value
$\alpha_s(M_Z)=0.118 \pm 0.005 $.

Our estimate of the NNLO theoretical uncertainties is based on application
of the [0/2] Pad\'e approach at the N$^3$LO level. The uncertainties
of our NNLO analysis can be decreased after explicit NNLO calculations of
the NS Altarelli-Parisi kernel.

As to the twist-4 terms, we found that despite the qualitative agreement
with the IRR model prediction, at the NNLO level they have the tendency
to decrease and are stable to
the application of the [0/2] Pad\'e motivated N$^3$LO analysis.

This feature can be related to the fact that
the analysis of the CCFR data can not distinguish
the twist-4 $1/Q^2$ terms
from the
NNLO perurbative QCD approximations of the Mellin moments.
This possible explanation
lies in the lines of the results of the LO analysis of the
old less precise neutrino DIS data, made by the authors of Ref.\cite{Barnett},
who were unable to distinguish between LO logarithmic and $1/Q^2$-behavior
of the QCD contributions to Mellin moments of $xF_3$.
The achieved in our days increase of experimental precision might move
this effect to the NNLO.

Another related explanation is  that the observed by us NNLO effect
is manifesting itself
in view of the fact that the detected by us  twist-4 terms
come from the partial summation of the definite terms of the
asymptotic perturbative QCD series and thus the increase of the order
of perturbative QCD analysis effectively suppresses the remaining sum
of the perturbative QCD contribution. Unfortunately, at the NNLO
level we can not detect the true twist-4 terms, which reflect the
non-trivial non-perturbative nature  of QCD vacuum.
One can hope that the future experiments of NuTeV
collaboration will allow to get the new experimental data at the
precision level, necessary for extracting more detailed information about
higher twist contributions to structure functions and will help to
clarify the reason of the disagreement of the low $x$ CCFR data for $F_2$
SF with the ones, obtained by the BCDMS collaboration. We hope
to return to the NNLO analysis of the experimental data of BCDMS
collaboration in the nearest future.

{\bf Acknowledgments}

The part of this work was done when one of us (AVS) was visiting
Santiago de Compostela Univ. He is grateful to his colleagues
for hospitality in Spain. Our studies were continued when ALK was
participating
at the LHC meeting at CERN (March, 1998). He is grateful to the Organizers
of this Meeting for the invitation and  support.

We are grateful to A.V.Kotikov for participation at the first stage
of these our studies and for discussions. The useful comments of
S.I. Alekhin  are gratefully acknowledged.
Special thanks are due to S.A.Kulagin for sharing his
points of view on the currents status of the studies of
the nuclear effects in DIS.

The preliminary versions of this work were reported
at Quarks-98 Int. Seminar (Suzdal, May, 1998),
Small-$x$ physics and Light Front Dynamics in QCD Conference
(St.Petersburg, July, 1998) and XI Int. Conference on Problems
of Quantum Field Theory in memory
of D.I. Blokhintsev (Dubna, July, 1998).
We would to thank the participants of these conferences, and especially
S.J. Brodsky, V.M. Braun, K.G. Chetyrkin, B.L. Ioffe, L.N. Lipatov, E.A. Kuraev
and A.V. Radyushkin  for the interest in this work.

\newpage

This work is supported by the Russian Fund for Fundamental Research,
Grant N 96-02-18897. The work of G.P. was supported by CICYT
(Grant N AEN96-1773) and Xunta de Galicia (Grant N XUGA-20604A96).

\newpage


\newpage
%

\end{document}